\pdfoutput=1
\documentclass[5p]{elsarticle}
\usepackage{hyperref}

\usepackage{graphicx}
\usepackage{amssymb, amsmath}

\usepackage{natbib}

\biboptions{round, semicolon, authoryear}

\newcommand{\dpm}[2]{\substack{+#2 \\ -#1}}

\begin{document}

\begin{frontmatter}


\title{Automated crater detection on Mars using deep learning}



\author{Christopher Lee}

\address{Toronto, Ontario, Canada}

\begin{abstract}
Impact crater cataloging is an important tool in the study of the geological history of planetary bodies in the Solar System, including dating of surface features and geologic mapping of surface processes. Catalogs of impact craters have been created by a diverse set of methods over many decades, including using visible or near infra-red imagery and digital terrain models.

I present an automated system for crater detection and cataloging using a digital terrain model (DTM) of Mars --- In the algorithm craters are first identified as rings or disks on samples of the DTM image using a convolutional neural network with a UNET architecture, and the location and size of the features are determined using a circle matching algorithm. I describe the crater detection algorithm (CDA) and compare its performance relative to an existing crater dataset. I further examine craters missed by the CDA as well as potential new craters found by the algorithm. I show that the CDA can find three--quarters of the resolvable craters in the Mars DTMs, with a median difference of 5-10\% in crater diameter compared to an existing database.

A version of this CDA has been used to process DTM data from the Moon and Mercury \citep{Silburt2019}. The source code for the complete CDA is available at \url{https://github.com/silburt/DeepMoon}, and Martian crater datasets generated using this CDA are available at \url{https://doi.org/10.5683/SP2/MDKPC8}.

\end{abstract}


\end{frontmatter}

\section{Introduction}
\label{S:introduction}

Information on crater populations and spatial distributions provide important constraints on the geological history of planetary surfaces. Regional differences in crater distributions and population characteristics can be used to constrain geologic processes and stratigraphy \citep{Cintala1976,Wise1980,Barlow2003,Barlow2005}, and crater populations can be used to estimate the age of surface features as well as constrain the timescale of surface processes \citep{Arvidson1974, Soderblom1974, Craddock1997, Stepinski2009b,Tanaka2014}. To enable such research, impact craters need to be identified, measured, and counted using imagery of a planet's surface.

However, the task of creating a dataset of crater locations has traditionally been a time--consuming process of manually identifying craters in printed maps \citep{Barlow1988} or digital imagery \citep{Robbins2012}. Recent advances in the quality of remote observations, in image processing techniques, and available computational power have lead to the development of advanced automated Crater Detection Algorithms \citep[CDAs, e.g.,][]{Stepinski2009, Di2014, Pedrosa2017, Silburt2019}. These CDAs use different approaches and datasets, but each method attempts to identify crater--like features on the surface using digital imagery or elevation datasets and apply image processing techniques to isolate the crater features. An important benefit of these CDAs is that it becomes possible to automate large parts of the crater--finding process and reduces the effort required after the initial implementation.

Automated CDAs have tunable parameters that can be optimized for the imagery or elevation dataset being processed. In designing the algorithms, a curated list of crater locations and images are used in a ``training'' step to adjust these parameters. Once trained, the CDA can be applied to larger datasets from the same body or even applied to different planetary bodies through “transfer learning” \citep{Silburt2019}.

In this work, I use an automated CDA based on a Convolutional Neural Network \citep[CNN,][]{Goodfellow2016} to identify \textit{circular} crater--like features in a martian Digital Terrain Model (DTM). I perform three experiments with the CDA to characterize its performance on the DTM under various assumptions. In two of the experiments, I attempt to find rings associated with the crater rims as in \citet{Silburt2019} using CNNs trained on lunar data (the \citet{Silburt2019} CDA) and martian data. In the third experiment, I train the CNN to find \textit{disk} structures associated with the entire crater. The latter method is commonly used in image segmentation methods to isolate features (e.g., cancerous cells in medical images,  \citet{Ronneberger2015}) and a similar method was developed by \citet{Stepinski2009}.

The CNN used in this work uses a standard ``UNET'' architecture \citep{Ronneberger2015} that is commonly used in image segmentation and processing, and similar CNNs have been applied to identification of tumors in medical images \citep{Cicek2016}, identification of radio frequency interference in astronomical data \citep{Akeret2017}, and crater detection on the Moon using a DTM from the Lunar Reconnaissance Orbiter \citep{Silburt2019}. The architecture of the CNN is not the primary purpose of this work, and the reader is referred to the referenced work for an in--depth discussion of the methodology.

The remainder of this paper is organized as follows: in section \ref{S:priorwork} I review prior work in developing automated CDAs; in section \ref{S:methods} I describe the crater detection algorithm, the training and data processing involved, and the structure of the experiments; in section \ref{S:results} I discuss the metrics calculated for each experiment and examine the new crater catalogs in detail; finally, in section \ref{S:conclusions} I provide concluding remarks.

The software used to make CDA described here is based on the work of \citet{Silburt2019} with three modifications: The source data used here retains the 16--bit raw precision of the source DTM compared to 8--bit image used in \citet{Silburt2019}; a disk--finding CNN is implemented with an additional processing step, discussed later; the distance and size thresholds used to determine duplicates and matches in the database were reduced to 0.25 of the crater diameter (from 2.6 and 1.8 diameter units).

The original code is available at \url{https://github.com/silburt/DeepMoon.git} , updates and modifications to the code can be found at \url{https://github.com/eelsirhc/DeepMars.git} . The catalogs generated here, along with ancillary data, can be found at \url{https://doi.org/10.5683/SP2/MDKPC8}. Similarly to \citet{Silburt2019}, I use Keras \citep{Chollet2015} version 2 with Tensorflow \citep{Abadi2016} to build, train, and test the model. In training the model I used an Nvidia 1080 Ti GPU using the CUDA and CUDNN support libraries, but the code is compatible with Intel and AMD CPUs.

In the results presented here, no manual corrections have been applied to the generated catalogs to improve the accuracy of the crater catalog. As a result, the CDAs miss around 25\% of the craters in the \citet{Robbins2012} dataset, and find an additional 25\% that do not match craters in the \citet{Robbins2012}. These numbers are significantly better than previous automated algorithms \citep[e.g.,][]{Stepinski2011}) and worse than current alternative crater catalogs created by expert human classifiers \citep{Salamuniccar2012}. While it would be possible to manually inspect the roughly 15,000 `new' crater candidates, or add the roughly 15,000 missed craters to the new catalog, that would not lead to an improvement in the performance of the CDA on observations that have not been studied by expert human classifiers. The better approach to improving the CDA on unseen data, tested here and in \citet{Lee2018a}, is to combine the results from multiple automated CDAs in a weighted `voting' algorithm (based on skill, as is done for human classifiers \citet{Robbins2017}) to provide a more robust automated method for cataloging craters.

\section{Prior Work}
\label{S:priorwork}

One of the first large global databases for Mars was created by \citet{Barlow1988} using printed maps from Viking orbiters and included 25,826 craters with a diameter greater than 8km. This dataset has been updated since then \citep{Barlow2003a} with 42,283 craters, and other datasets are available (\citet{Rodionova2000} with 19,308 craters, \citet{Salamuniccar2008} with 57,633 craters). The most comprehensive dataset for Mars craters is that derived from the Thermal Emission Imaging System instrument by \citet{Robbins2012}. The \citet{Robbins2012} dataset includes 383,343 craters with diameters greater than 1km, including 30,473 craters above 8km diameter. These craters were identified in 256 pixel/-degree resolution THEMIS IR imagery using a customized manual image processing pipeline. The \citet{Robbins2012} dataset is reported to be statistically complete to 1km diameter for the majority of Mars covered by the source THEMIS dataset, reflected in the power law distribution following the expected distribution to diameters of 1km or lower \citep{Arvidson1979}. In this work, I consider craters with a diameter greater than 4km based on the resolution limit of the input DTM.

In contrast to the attempts to catalog martian craters using manual methods, automated CDAs have not been extensively used to generate \textit{global} catalogs of martian craters, but have been used to catalog small test areas containing a mixture of crater types.  Two common metrics used in machine learning comparisons are \textit{precision} and \textit{recall}, whose mathematical definitions are given in the next section. \textit{Precision} is the fraction of craters found by the CDA that exist in a target dataset (usually a subset of \citet{Barlow1988} or \citet{Robbins2012}), and the \textit{recall} is the fraction of craters in the target dataset that are found by the CDA. In the calculation of precision, the effect of detected craters that are real, but do not exist in the target dataset, is to decrease the precision.

\citet{Stepinski2009} developed the AutoCrat CDA using the 128pixels\slash degree DTM from the Mars Orbiting Laser Altimeter (MOLA) instrument. The AutoCrat CDA combines a ``rule-based'' module that applies gradient--based algorithms to identify local depressions as possible craters, followed by a ``machine-learning'' module that applies a decision tree algorithm \citep{witten2005} to determine whether the feature is a crater or not. The decision tree algorithm is used to differentiate craters from non--crater depressions using diameter, depth, and shape parameters as factors. Only a small fraction of the planet is covered in \citet{Stepinski2009}, with a reported precision of 42\% (1,544 known craters out of 3,666 detections) and a recall of 72\% (1,544 out of 2,144 known craters were found). A global database generated using this CDA was reported in \citet{Stepinski2009b} with 75,919 craters larger than 1.37km.

\citet{Di2014} developed a CDA that also processed DTM images. Their CDA uses a sliding window correlator to find and highlight crater edge features and a circular Hough transform to transform those highlighted crater edges into locations and sizes. \citet{Di2014} reports on the CDA performance for three sites with 11,868 craters, but they do not provide an explicit calculation of precision and recall. \citet{Di2014} reports finding 934 craters in total, with 114 false positives (a precision of 87\%) with a recall rate using the same data (their table 2) of 74\% for craters with a diameter greater than 6km, but a recall rate of less than 10\% for all craters tested. 

\citet{Pedrosa2017} developed a CDA using thermal imagery from THEMIS. The CDA processes THEMIS IR imagery by first identifying geophysical depressions using a `watershed' transform (to find virtual floodplains) and then within each watershed identified the local minima as possible craters. A circle template matching algorithm is then used to compare the crater features to a characteristic ring representing the crater rim. \citet{Pedrosa2017} reports a precision and recall of 65\% and 91\%, respectively (their figure 7), compared to a target dataset of 3,600 craters. The template matching system employed by \citet{Pedrosa2017} is similar to the method used in \citet{Silburt2019} and here to find the location of each crater.

\citet{Salamuniccar2011} provides a summary of many more automated CDAs, and a discussion of the effect of combining multiple CDAs into one dataset. The various methods used in the CDAs are essentially the same as the CDA developed here. A correlation function is used to identify features that identify a crater --- edges in \citet{Di2014}, disks in \citet{Stepinski2009}, opposing crescents in \citet{Pedrosa2017}. Once the crater is identified a circle finding algorithm is used to determine the location and size --- a Hough transform in \citet{Di2014}, a sliding window correlator in \citet{Pedrosa2017}. In the CDA developed here, discussed in detail in the next section, the CNN implements a sequence of correlation functions to identify and highlight the crater, followed by a circle matching algorithm to determine the location of the craters.

\section{Methods}
\label{S:methods}
\subsection{Input dataset}
The source digital terrain model (DTM) for this work is the “Mars HRSC MOLA Blended DEM Global 200m v2” \citep{Fergason2018} dataset available from the Astrogeology Science Center website (https:\slash \slash astrogeology.usgs.gov). This map is a blend of digital terrain models derived from the Mars Orbiter Laser Altimeter \citep[MOLA, ][]{Smith2001} aboard the Mars Global Surveyor spacecraft, and the High-Resolution Stereo Camera \citep[HRSC, ][]{Jaumann2007} aboard Mars Express. The stated scale of the dataset is 200m\slash pixel horizontally, chosen as a compromise between the 463m\slash pixel scale of MOLA and the 50m\slash pixel scale of HRSC. However, the HRSC DTMs that were included cover only 44\% of the planet, so more than half of the planet is interpolated from MOLA dataset at 463m\slash pixel scale. Some regions have no data from either spacecraft. The stated accuracy of each point is 100m horizontally and at best 1m radially \citep{Fergason2018}. The total image size of the DTM is $106694\times\,53347$ pixels with 16--bit resolution for the elevation data, using a simple cylindrical (Plate Carr\'ee) projection. The effective resolution of this source image is $\frac{1}{296}$\textsuperscript{th} of a degree, and $\frac{1}{2}$ m vertically (better than the resolution of the input data).

The CDA takes as input a $256\times\,256$ pixel 8--bit image taken from the larger DTM and attempts to identify craters within this image. To prepare an image for use with the CDA I use the following steps:

\begin{enumerate}
	\item A square sample is extracted from the DTM and resampled into the required $256\times\,256$ size.\label{pre:step1}
	\item The bit resolution of the image is rescaled from the 16--bit source to the 8--bit resolution required for the CDA. This step occurs after resampling the image to a smaller region to mitigate the effect of the large altitude variation on Mars.\label{pre:step2}
	\item The image is orthographically projected using the Cartopy Python package \citep{Cartopy} to provide an image with near--constant linear scale instead of the constant angular resolution of the Plate Carrée projection.\label{pre:step3}
	\item Padding is added to the image as required to fill in the square image after projection. The Orthographically projected image always occupies fewer pixels than the Plate Carrée source image.\label{pre:step4}
\end{enumerate}

The size of the image sample (step \ref{pre:step1}) is chosen from a list of sizes from 512 to 16,384 pixels to provide a range of scales from 400m/pixel to 12.8km/pixel. The full dataset is constructed by sampling the entire planet at a range of pixel scales and with overlapping regions between adjacent images, requiring 55,000 images in total. I also tested an additional 150,000 images sampled at the original scale of the DTM (200m\slash pixel), but the performance of the CDA degrades substantially because of the coarser scale of the majority of the input DTM.  An alternate method used by \citet{Silburt2019} was to select the location and size of the images at random, which provides similar statistical results to the systematic method above, but would not guarantee planet-wide coverage. 

In the current implementation of the CNN each image takes 0.5Mbit of storage, but this is not the limiting factor in the algorithm. Each instance of the UNET CNN requires storing 10 million parameters (the connections between each layer) and uses 600MB of storage. During training 16 images are simultaneously processed requiring 9.6GB of memory. At this scale, a consumer video card (Nvidia 1080 Ti) is approximately an order of magnitude faster than a 32 core Intel Xeon workstation with 192GB of memory. Further, the GPU is optimized for 8--bit integer operations compared to 16--bit floating point operations, and can process an 8--bit image up to 200 faster than the equivalent 16--bit image.

The source crater catalog for this work is the \citet{Robbins2012} catalog, with 383,343 craters larger than 1km in diameter. To mitigate problems with the resolution of the DTM (discussed above) and possible problems with craters smaller than 2km in diameter from the \citet{Robbins2012} catalog \citep{Robbins2017} I include only craters larger than 4km in diameter in this work.

\subsection{Experiments}
\label{SS:experiments}
I performed three experiments with the CDA. The first experiment uses a CNN trained on Lunar data \citep{Silburt2019} to find ring structures associated with the crater rim. This CNN has not been previously trained on Mars crater observations and is an example of \textit{transfer learning}.

The second experiment modifies the first by training the CNN using a subset of the Mars crater imagery without using the previously Moon trained CNN. The target data for the training is derived from a human--generated crater database \citep{Robbins2012} using high--resolution infra--red imagery. In the second experiment, the CNN is trained to identify rings associated with the rims of craters, and a summary of the training method is provided in the next subsection.

In the third experiment, the CNN is modified to identify \textit{disks} associated with craters, instead of rings. This approach follows the same training methodology as the ring finding CNN, but with a modified training dataset and crater matching algorithm. To keep the comparison as close as possible I use the same image locations for both trained CNNs, and in the \textit{disk} finding CDA I convert crater features highlighted by the CNN to ring structures before attempting to locate the craters. This disk--ring conversion makes comparison easier between the CNNs but is not necessarily an optimized algorithm for the disk finding CDA.

\subsection{Training and Validation}
\label{SS:training}
\begin{figure*}
\centering
\includegraphics[width=6in]{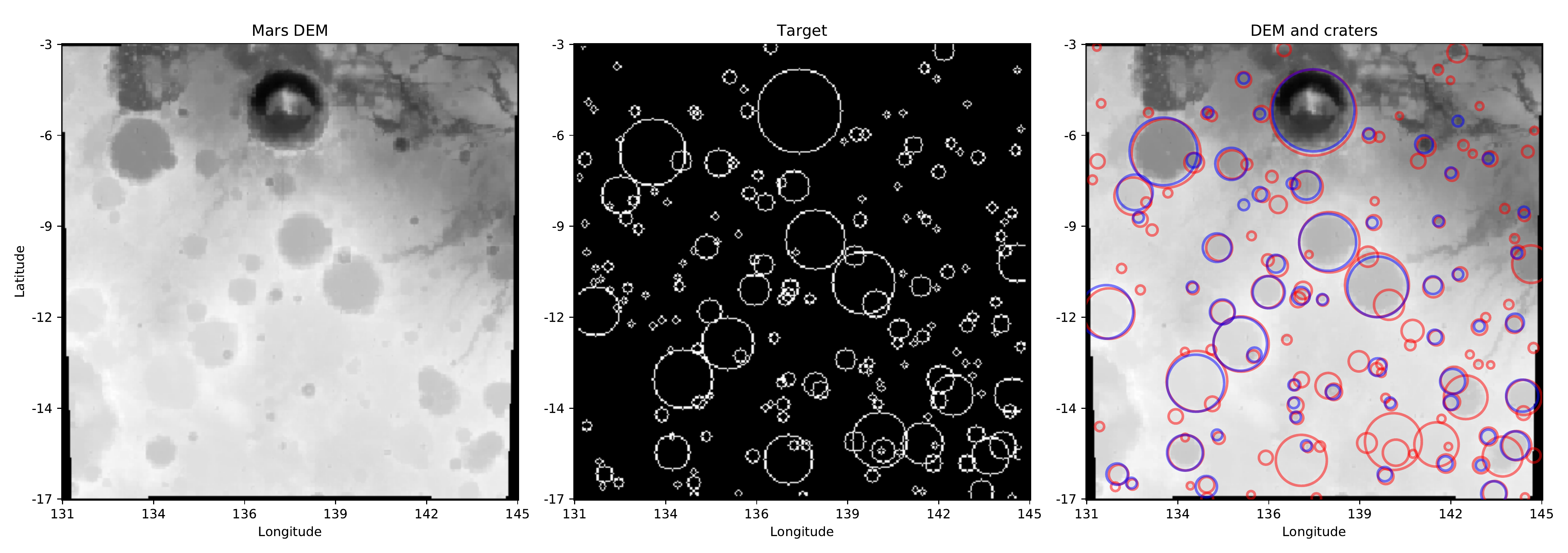}
\caption{Example DTM image (left),target map (center), and identified craters (right). The DTM is extracted from the source HRSC+MOLA map at 137E longitude, 8S latitude with a resolution of 18 pixels per degree (approximately 3.2 km\slash pixel). Gale crater is located at the center--top of the image, and is found by the CNN in this example. The right panel has a 4 pixel diameter cutoff and craters near this threshold might appear in one catalog but not the other simply because of this strict cutoff. Red circles show craters from the \citet{Robbins2012} dataset, blue circles show craters found by the CDA.}\label{F:example}
\end{figure*}

Training and validation follows the method outlined in \citet{Silburt2019}, and an example is given in figure \ref{F:example}. A sample image is taken from the dataset (figure \ref{F:example}--left) and the locations of resolvable known craters are taken from the \citet{Robbins2012} and drawn as white pixels on an otherwise black `image' (figure \ref{F:example}--center).  The CNN is then trained by exposure to a large number of images and trained to encode the DTM into the binary ring image (figure \ref{F:example}--center). Figure \ref{F:example} is one of the 15 images in the dataset that includes a majority of Gale crater in the image -- this figure is centered on 137 degrees east longitude, 8 degrees south longitude, at 3.2km/pixel scale. 

Figure \ref{F:example}--right super-poses the input DTM image with the known craters \citep{Robbins2012} in red and craters found by the ring finding Mars--trained CDA in blue. In this image overlapping blue and red circles identify craters correctly identified by the CNN (though some are displaced spatially), red circles with no blue counterpart are missed craters (\textit{false negatives} in machine learning terminology), while blue circles with no red counterpart are features incorrectly identified as craters by the CNN (\textit{false positives}). In principle, the false positives might be new craters, but I will suggest later that the majority are not new craters, although some are genuine circular features (e.g., paterae). The rightmost panel of figure \ref{F:example} has a strict 4--pixel diameter cutoff such that craters might be removed from one catalog and not the other, appearing as false positives (only blue circles) or false negatives/missed craters (only red circles) even if a matching smaller crater exists.

For the CNN trained with martian craters, I use 30,000 images distributed in location and scale in the training dataset (5,000 images were reserved for a testing stage during the training), and 25,000 images in the validation dataset. The images are distributed geographically so that both datasets contain unique images but similar spatial distributions. The image extents are also distributed between datasets, with similar numbers of each scale in each dataset. The small number of large geographically extended images (fewer than 500) means they are unevenly distributed between the two datasets.

A single iteration of the training uses all 20,000 training images and measures the accuracy of the updated CNN using the 5,000 reserved test images. This training is repeated until the test accuracy stops improving, or after 30 iterations. In \citet{Silburt2019}, the CNN was trained with 30,000 images of the Moon for 4 complete iterations (120,000 images in total). For the Mars trained CNN the test accuracy stopped improving after 8--10 iterations (200,000 to 250,000 images). The difference in total number of images used does not necessarily translate to an improved CDA --- The largest improvements occur early in the training process, and \citet{Silburt2019} also used a larger 30,000 image test dataset that would have smaller statistical fluctuations in the accuracy metrics.

The CNN does not produce a location or size for each crater in the image but instead transforms the DTM image into a binary image that highlights topographic features that are related to craters. The CNN is best at highlighting features that are between 10 and 60 pixels in diameter in any particular image, resulting in a large number of missing craters in each image. Small craters are represented by too few pixels  to be positively identified. Large craters become diffuse or incomplete circles fall below the detection threshold. In figure \ref{F:example}, Gale crater was among the largest identified features, even though a few larger craters are visible in the image.

\subsection{CNN Processing}
\label{SS:cnnprocessing}

The location and size of the craters in the CNN images are determined using the \texttt{match\_template} algorithm from scikit-image \citep{VanderWalt2014}. The \texttt{match\_template} algorithm finds the location and size of each circular feature by maximizing its correlation with a \textit{template circle} of known size. To allow comparison with the \citet{Silburt2019} study I keep the same threshold parameters for the circle matching algorithm, though small improvements may be possible with more extensive re--training of the algorithm. For each crater map generated by the CNN the location and size of craters are found with the following steps:

\begin{enumerate}
\item A candidate circle size is chosen and used to generate a template image.\label{post:step1}
\item The candidate circle is compared against all locations in the binary crater image generated by the CNN. The resulting map becomes a ``heat map'' of correlation between the candidate circle and the template.\label{post:step2}
\item Where the correlation between the crater map and the candidate circle exceeds a confidence threshold that location is identified as the crater location and the size of the candidate circle is used as a crater size.\label{post:step3}
\end{enumerate}
This template matching process is conducted for circles with integer radii from 5 pixels to 40 pixels. For architectural reasons the CNN rarely predicts craters smaller than 10 pixels in diameter or larger than 60 pixels, with typical minimum and maximum diameters of 10 pixels and 30 pixels, respectively. The circle matching algorithm performs poorly for circles with a diameter smaller than 10 pixels where it considers diffuse segments of larger craters as potential small craters, resulting in 'rings' of small craters around larger craters. Duplicate craters are removed in each image by identifying craters that are within 0.25 diameter units in size and within 0.25 diameter units in location of another crater. These values are smaller than those used by \citet{Silburt2019}.

The result of this post-processing is a list of unique craters found in each input image, before any comparison with the \citet{Robbins2012} database. In the right panel of figure \ref{F:example} this post-processing produced the blue circles, while the \citet{Robbins2012} craters with a diameter of at least 4 pixels are shown as red circles.

The disk--finding CNN is trained to highlight craters by replacing the crater in the DTM with a solid disk in the binary image, instead of a ring surrounding the crater. After the disk CNN has processed the DTM scene a \citet{Sobel1968} transform is used to convert this disk into a ring to emulate the output of the ring--finding CNN. After this additional step, the analysis follows the ring finding method above.

A potential downside of the disk finding method is that overlapping craters are not easily separated. Overlapping craters found by the disk finding CNN are typically represented as non circular features and therefore are rejected by the circle matching algorithm. One possible improvement used by \citet{Stepinski2009} is to use a pre--processing step to filter the images for preferred length scales using a Gaussian blurring technique, this would better separate large and small craters regardless of their overlap, but would not separate similarly sized overlapping craters. The disk finding algorithm is not recommended as the only CDA for this reason.

\subsection{Post Processing}
\label{SS:postprocessing}

The image dataset contains 55,000 images with resolutions ranging from 150 pixels/degree to 4 pixels/degree covering the planet. As a result, a single location would appear at up to 7 different resolutions in $15\pm6$ images (the variation is due to the use of overlapping images). For example, Gale crater appears at 5 resolutions in 9 images.

\begin{figure*}
\centering
\includegraphics[keepaspectratio,width=0.8\textwidth]{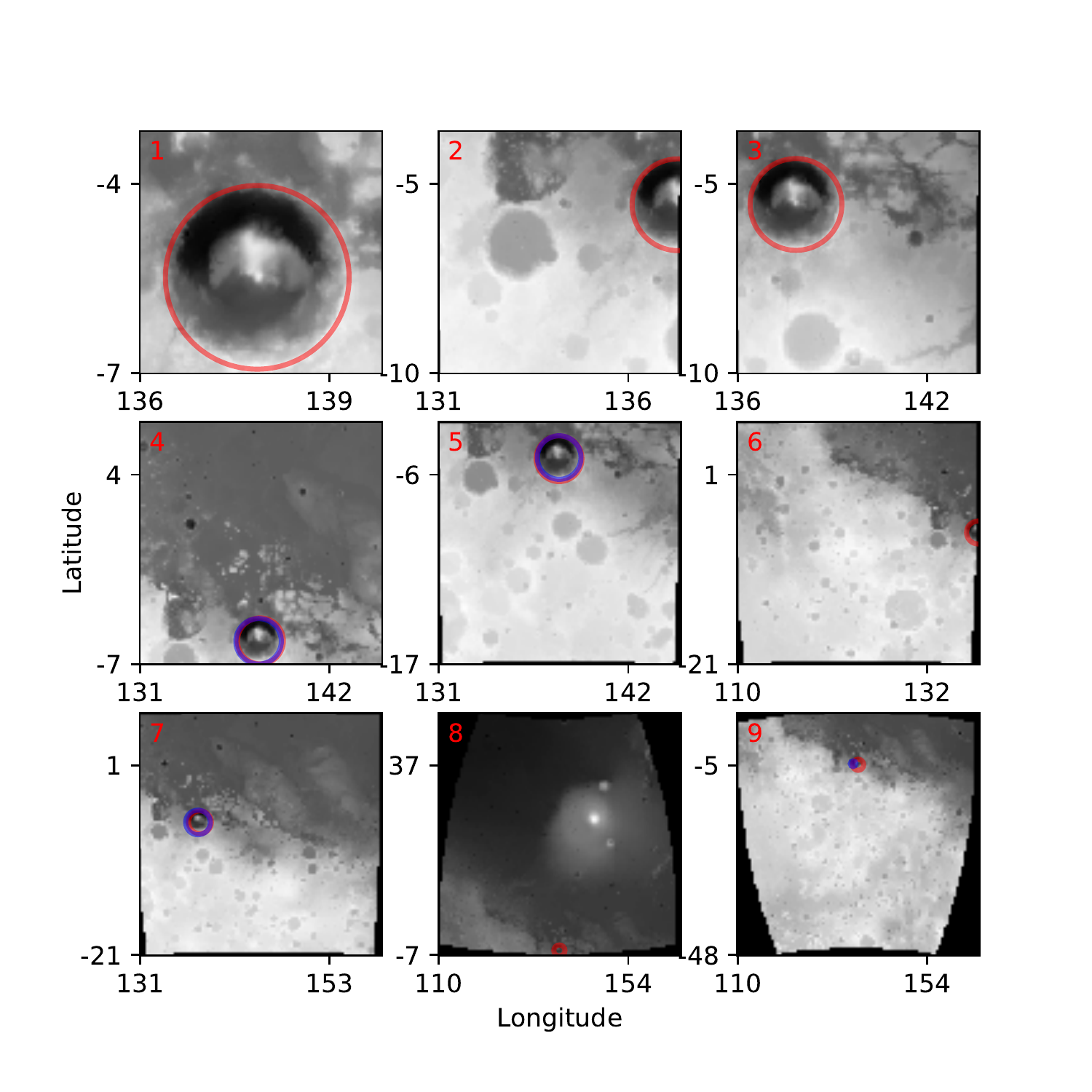}
\caption{DTM images where more than 50\% of Gale crater is contained in the image. Gale crater is highlighted with a red circle (using the \citet{Robbins2012} location) and a blue circle where it was identified by the CDA in each image. The CDA identified the crater in 5 images in this example data (4,5,7,8,9). The crater in images 1,2, and 3 is probably too big for the current algorithm. Image 2 and 6 only include partial circles and lie below the detection threshold.}\label{F:example_gale}
\end{figure*}

Each of the 55,000 images is processed by the CNN and template matching algorithm to find craters independently of the other images. The location and size of each crater is found in pixel space during the template matching stage and then converted into geographic coordinates using the known limits of each image and the orthographic projection parameters.

As a result of the overlapping images and multiple resolutions, single craters may be identified in multiple images and appear multiple times in the generated global crater list. Duplicate craters are removed by comparing the diameter and location of the crater with other craters with similar location and size, using the same parameters as in the last section.

Figure \ref{F:example_gale} shows the 9 images that include more than 50\% of Gale crater. In this example, Gale crater was found in 5 images. The 5 candidate craters are combined in the post-processing stage to provide 1 location and size for Gale crater, preferentially using the values found in the highest resolution image.

Once all duplicates are removed the final result is a database with approximately 60,000 craters found by the CDA. This list includes only craters larger than 4km in diameter, the lower limit allowed in the algorithm. Above this 4km limit, the ring CDAs found 75\% of all craters listed in the \citet{Robbins2012} database. Above 10km diameter, the ring CDAs found more than 80\% of all craters in the \citet{Robbins2012} database. The algorithm itself has no lower limit in geophysical space but does have a lower limit in pixel space. The circle--matching algorithm works well for circles 10 pixels in diameter or larger and continues to work down to 6--pixel diameter circles but with a higher false positive rate. Ten pixels in diameter represents a physical limit of 4km in diameter using the 400m/pixel scale images. As a comparison, \citet{Robbins2012} used 256 pixel/degree THEMIS infra-red imagery that covers more of the planet than the DTM used here and the crater size limit reported in \citet{Robbins2012} is 1km, corresponding to an absolute minimum of 5 pixels in diameter at the equator. \citet{Robbins2013} suggested a lower diameter limit of 10km for MOLA derived DTM data, noting that the higher resolution DTMs derived from HRSC are better at resolving smaller craters.

\subsection{Accuracy Metrics}
\label{SS:accuracymetrics}
In each experiment, the performance of the CDA is measured against the \citet{Robbins2012} database using standard metrics. The crater locations in the CDA database and the \citet{Robbins2012} database are compared using the same methodology used to find duplicate craters. If the CDA finds a crater within 0.25 diameter units in location and size of a crater from the \citet{Robbins2012} database then it is considered a match. A \textit{True Positive} is a match between the CNN and \citet{Robbins2012} database, a \textit{False Positive} is a crater in the CDA database without a matching crater in the \citet{Robbins2012} database, and a \textit{False Negative} is a crater in the \citet{Robbins2012} database without a matching crater in the CDA database. \textit{True Negatives} are not used.

Using these definitions, the precision $P$ is defined as the ratio of true positive to all identifications, and the recall $R$ as the ratio of true positives to all craters in the \citet{Robbins2012} database.

\begin{eqnarray}
P&=& \frac{T_p}{T_p + F_p},\\
R&=& \frac{T_p}{T_p + F_n}
\end{eqnarray}

Where $T_p$, $F_p$, and $F_n$ are the numbers of true positives, false positives, and false negatives, respectively. A high precision suggests the CDA has a high fractional true positive rate, while a high recall suggests the CDA finds a high fraction of the existing craters.

As an extreme example, the CDA could identify craters everywhere on Mars whether they exist or not, resulting in a perfect recall (all craters are found) but almost no precision (many false positives are found). Alternatively, the CDA could identify a single crater correctly, resulting in perfect precision (no false positives) but almost no recall (many missing craters, or having many false negatives). A common metric used to balance the precision and recall is the harmonic average of the two metrics, commonly called the $F_1$ score, 

\begin{equation}
F_1 = \frac{2PR}{P+R}
\end{equation}
where the same $F_1$ value can be found using different combinations of precision and recall. None of these metrics reward identification of new craters that do not exist in the \citet{Robbins2012} database. All new identifications are assumed false positives and reduce the precision and $F_1$ score. The possible new crater fraction N is calculated as the ratio of false positives to the sum of false positive and \citet{Robbins2012} craters,

\begin{equation}
N = \frac{F_P}{F_P + T_R}
\end{equation}
Where $T_R$ is the number of true craters in the \citet{Robbins2012} database. This is an upper limit on the number of new craters found, and because the \citet{Robbins2012} database is statistically complete below the 4km limit used here it is likely that many false positive are genuine false positives and not new craters. In \citet{Silburt2019} a sample of the false positive craters was studied and they estimated that 89\% of the false positive craters are new craters (11\% are genuine false positives, \citet{Silburt2019} table 1).

For further comparison with the \citet{Robbins2012} database I also calculated the difference in longitude, latitude, and diameter between the CNN craters and \citet{Robbins2012} craters, both in pixel units and geophysical units. Each of these metrics is calculated as the ratio relative to the smallest crater diameter in the comparison and given as the mean value and interquartile ranges for the dataset.

\subsection{Error Sources}
\label{SS:errorsources}
A few sources of error are present in the experiments, from observational constraints, pixelization of the source data, and the algorithm design.

The source DTM combines HRSC and MOLA data at a stated scale of 200m/pixel. However, this is obtained by upsampling the MOLA data from 463m/pixel for 56\% of the surface, and downsampling HRSC images for the remaining 44\%. The smallest image scale used in this experiment was 400m\slash pixel, close to the MOLA laser footprint of 300m, giving a accuracy limit of .4km in crater location and diameter (i.e., 1 pixel).

The crater position is extracted by matching circular templates on images. The discrete nature of the images limits the matches to 1 pixel in any image. At the highest image resolution used this corresponds to the .4km accuracy limit above, but in images with a larger pixel scale the accuracy decreases at a corresponding rate. At the 12.8km/pixel scale for the largest images, the accuracy drops to about 6km at the equator. In practical terms, a crater is likely to be found when it is between 10 pixels and 30 pixels in diameter, making the 1 pixel uncertainty equivalent to a 3--10\% error in position or diameter. When the global CDA database is generated, the highest resolution image that included the crater detection was used to obtain the best overall position and size data for each crater in the final dataset. For example, in the Gale crater example in figure \ref{F:example_gale}, image 4 or 5 would be used when calculating the location and size of the crater.

Projecting the DTM from Plate Carée into orthographic and back introduces some errors depending on the extent of the image, as distortion increases away from the center point of the projection. \citet{Silburt2019} estimated an error of 2\% in the crater size for typical images, which becomes larger than the pixelization errors for craters larger than about 50 pixels in diameter. Few craters were found larger than 30 pixels in diameter so the contribution of this error is negligible.

Finally, algorithmic implementation also introduces some uncertainty. In the CNN step, the image bit resolution is limited to 8--bits of data, which for large images with vertically extended topography would obfuscate shallower craters -- 1km of vertical extent requires a vertical resolution of 4m at best, while an image that includes Olympus Mons and the surrounding terrain might be limited to 100m vertical resolution. In the template matching step, spurious matches can occur when comparing small candidate craters against large craters as the template matches along the crater wall, or comparing candidate circles against the space \textit{between} two or more nearby craters if the 'void' between the craters can be identified as a crater. 

\section{Results}
\label{S:results}

\begin{table*}[h]
	\centering
	\scalebox{0.7}{
	\begin{tabular}{|c|cc||cc||cc|}
		\hline
								& \textbf{Moon} 	& \textbf{Moon} 	& \textbf{Mars} 	& \textbf{Mars} 	& \textbf{Disk} 	& \textbf{Disk}     \\
								& \textbf{(image)}  & \textbf{(global)} & \textbf{(image)} 	& \textbf{(global)} & \textbf{(image)}  & \textbf{(global)} \\
		\hline
		Crater count		 	& $9.9\pm10.0$		& $57,564$ 			& $9.9\pm10.0$		& $57,564$ 			& $9.9\pm10.0$		& $57,564$ \\ 
		Craters detected		& $4.8\pm5.1$		& $54,739$ 			& $4.9\pm5.2$		& $57,767$ 			& $5.1\pm4.9$		& $75,733$ \\
		Craters matched		 	& $4.3\pm4.7$		& $42,445$ 			& $4.4\pm4.8$		& $42,891$ 			& $4.3\pm4.7$		& $39,149$ \\
		Latitude Error		 	& $4\dpm{1}{2}$ 	& $2\dpm{1}{1}$ 	& $4\dpm{2}{2}$ 	& $2\dpm{1}{1}$ 	& $4\dpm{2}{2}$ 	& $2\dpm{1}{1}$ \\
		Longitude Error		 	& $5\dpm{2}{2}$ 	& $2\dpm{1}{2}$ 	& $5\dpm{2}{2}$ 	& $3\dpm{1}{2}$ 	& $5\dpm{2}{2}$ 	& $2\dpm{1}{2}$ \\
		Diameter Error		 	& $6\dpm{3}{3}$ 	& $5\dpm{3}{4}$ 	& $7\dpm{3}{3}$ 	& $6\dpm{3}{4}$ 	& $9\dpm{4}{4}$ 	& $6\dpm{3}{5}$ \\
		Percentage new craters	& $5\pm8$			& $18$ 		 		& $5\pm8$			& $21$ 				& $7\pm11$			& $39$ \\ 
		Maximum diameter (pix)	& $34.1\pm20.0$		& $-$  		 		& $33.5\pm19.8$		& $-$  				& $32.7\pm18.9$		& $-$ \\ 
		Precision		 		& $90\pm18$			& $78$ 		 		& $89\pm19$		 	& $74$ 				& $84\pm23$		 	& $52$ \\ 
		Recall		 			& $42\pm21$			& $74$ 		 		& $43\pm21$		 	& $75$ 				& $44\pm23$		 	& $68$ \\
		F1		 				& $58\pm17$			& $76$ 		 		& $59\pm17$		 	& $74$ 				& $58\pm18$		 	& $59$ \\ 
		\hline
	\end{tabular}}
	\caption{Metrics calculated for three neural network based CDAs. ``Moon trained'' and ``Mars trained'' refer to the data used to train the initial model. ``Disk'' trained using Mars crater imagery in training, but identified disks associated with craters instead of crater ``rings''. All metrics were calculated using the same martian crater images from MOLA\slash HRSC. Values are given as mean $\pm$ 1 standard deviation (for single values after the $\pm$) or median $\pm$ inter--quartile range (for two values after the $\pm$) as in \citet{Silburt2019}. Precision, recall, and $F_1$ scores are given as percentages. Each model appears twice, with the ``image'' column given the per image metrics aggregated over the ensemble of 55,000 images (after removing duplicates per image), and the ``global'' column gives the post-processed metrics (after removing duplicates globally).}\label{T:global}
\end{table*}

Table \ref{T:global} gives the metrics for the three different experiments. All of the metrics are calculated using the same source images but the training of each network is different. The ``Moon'' trained CDA uses the network generated by \citet{Silburt2019} with no further training. The ``Mars'' trained CDA uses the network trained on a subset of the martian crater catalog, where the network is trained to find the crater rim. The ``Disk'' trained network is also trained on martian crater images, but is trained to find the whole disk of the crater, instead of just the rim. Table \ref{T:global} includes data from 55,000 images derived from the global MOLA\slash HRSC DTM. As a comparison, the same crater matching analysis was performed using the \citet{Salamuniccar2012} catalog in comparison to the \citet{Robbins2012} catalog. In this case, the recall is 79\% (21\% of the craters in the \citet{Robbins2012} catalog have no matching crater in the \citet{Salamuniccar2012} catalog) and the precision is 92\% (8\% of the craters in the \citet{Salamuniccar2012} catalog have no matching crater in the \citet{Robbins2012} catalog). Equivalently, the recall and precision of the \citet{Stepinski2011} crater catalog relative to \citet{Robbins2012} is 59\% and 50\%, respectively.

A smaller dataset using only images that were withheld during the training phase for the “Mars trained” CNN is shown in table \ref{T:validation}. None of the CDAs were shown the images summarized in table \ref{T:validation} during training.
\begin{table*}[h]
	\centering
	\scalebox{0.7}
	{\begin{tabular}{|c|cc||cc||cc|}
		\hline
								& \textbf{Moon} 	& \textbf{Moon} 	& \textbf{Mars} 	& \textbf{Mars} 	& \textbf{Disk}    & \textbf{Disk} 		\\
								& \textbf{(image)} 	& \textbf{(global)} & \textbf{(image)}  & \textbf{(global)} & \textbf{(image)} & \textbf{(global)}  \\
		\hline
		Crater count			& $8.1\pm5.3$		 & $32,979$ 		& $8.1\pm5.3$	    & $32,979$ 			& $8.1\pm5.3$		 & $32,979$ \\
		Craters detected		& $3.9\pm3.1$		 & $26,808$ 		& $4.0\pm3.2$	    & $28,198$ 			& $4.2\pm3.2$		 & $34,419$ \\
		Craters matched			& $3.4\pm2.9$		 & $21,732$ 		& $3.5\pm2.9$		& $21,985$ 			& $3.5\pm2.9$		 & $19,949$ \\
		Latitude Error			& $4\dpm{1}{2}$ 	 & $2\dpm{1}{1}$ 	& $4\dpm{2}{2}$		& $1\dpm{1}{1}$ 	& $4\dpm{2}{2}$ 	 & $1\dpm{1}{1}$ \\
		Longitude Error			& $5\dpm{2}{2}$ 	 & $2\dpm{1}{2}$ 	& $5\dpm{2}{2}$		& $2\dpm{1}{2}$ 	& $5\dpm{2}{3}$ 	 & $2\dpm{1}{2}$ \\
		Diameter Error			& $6\dpm{3}{3}$ 	 & $5\dpm{3}{4}$ 	& $7\dpm{3}{3}$		& $6\dpm{3}{4}$ 	& $8\dpm{4}{5}$ 	 & $6\dpm{3}{5}$ \\
		Percentage new craters	& $4\pm8$			 & $13$		 		& $5\pm8$		 	& $16$				& $8\pm11$		 	 & $30$ \\ 
		Maximum diameter (pix)	& $34.1\pm20.3$		 & $-$ 	 			& $33.6\pm20.2$		& $-$ 				& $32.6\pm19.1$		 & $-$ \\
		Precision		 		& $90\pm18$			 & $81$ 			& $89\pm19$		 	& $78$				& $83\pm24$			 & $58$ \\
		Recall		 			& $42\pm22$			 & $66$ 			& $43\pm22$		 	& $67$				& $44\pm23$			 & $60$ \\
		F1		 				& $58\pm18$			 & $73$		 		& $59\pm18$		 	& $72$				& $58\pm19$			 & $59$ \\ 
		\hline
	\end{tabular}}
	\caption{Metrics calculated using the validation dataset as in table 1, but for a subset of the images not used in training the ``Mars trained'' or ``Disk trained'' networks.}\label{T:validation}
\end{table*}

When training machine--learning algorithms there is a risk of `overfitting' where the algorithm becomes significantly better (by some metric) on the dataset it is trained with, at the expense of performing poorly on data it has not been shown. This overfitting can be seen when the precision (or recall, or F1 score) of the algorithm is much higher for a `training' dataset than an unseen `validation' dataset. Comparing the results in table \ref{T:global} and \ref{T:validation}, the metrics calculated for the validation dataset and the complete (validation and training) dataset suggests the networks are not overfitting the training data. This is reinforced by the performance of the ``Moon'' trained CDA that has never been trained using the Mars dataset. Differences between the global and validation metrics for this CDA reflect statistical differences in the performance of the CDA on the two datasets.


In the following subsections I examine the performance of the CDA in more detail. The results are separated by the type of detection: section \ref{SS:allcraters} examines all CDA crater detections in comparison to the \citet{Robbins2012} dataset; section \ref{SS:matchedcraters} examines at the matched (\textit{true positives}) in the CDA datasets; section \ref{SS:missedcraters} examines the craters missed by the CDA (\textit{false negatives}), and I use the extended data provided in the \citet{Robbins2012} dataset to identify the characteristics of the those missing craters; finally, section \ref{SS:falsepositives} examines the craters found by the CDAs that do not exist in the \citet{Robbins2012} dataset -- the false positives.

\subsection{All Craters}
\label{SS:allcraters}
\begin{figure*}[htbp]
	\centering
	\includegraphics[keepaspectratio,width=0.8\textwidth]{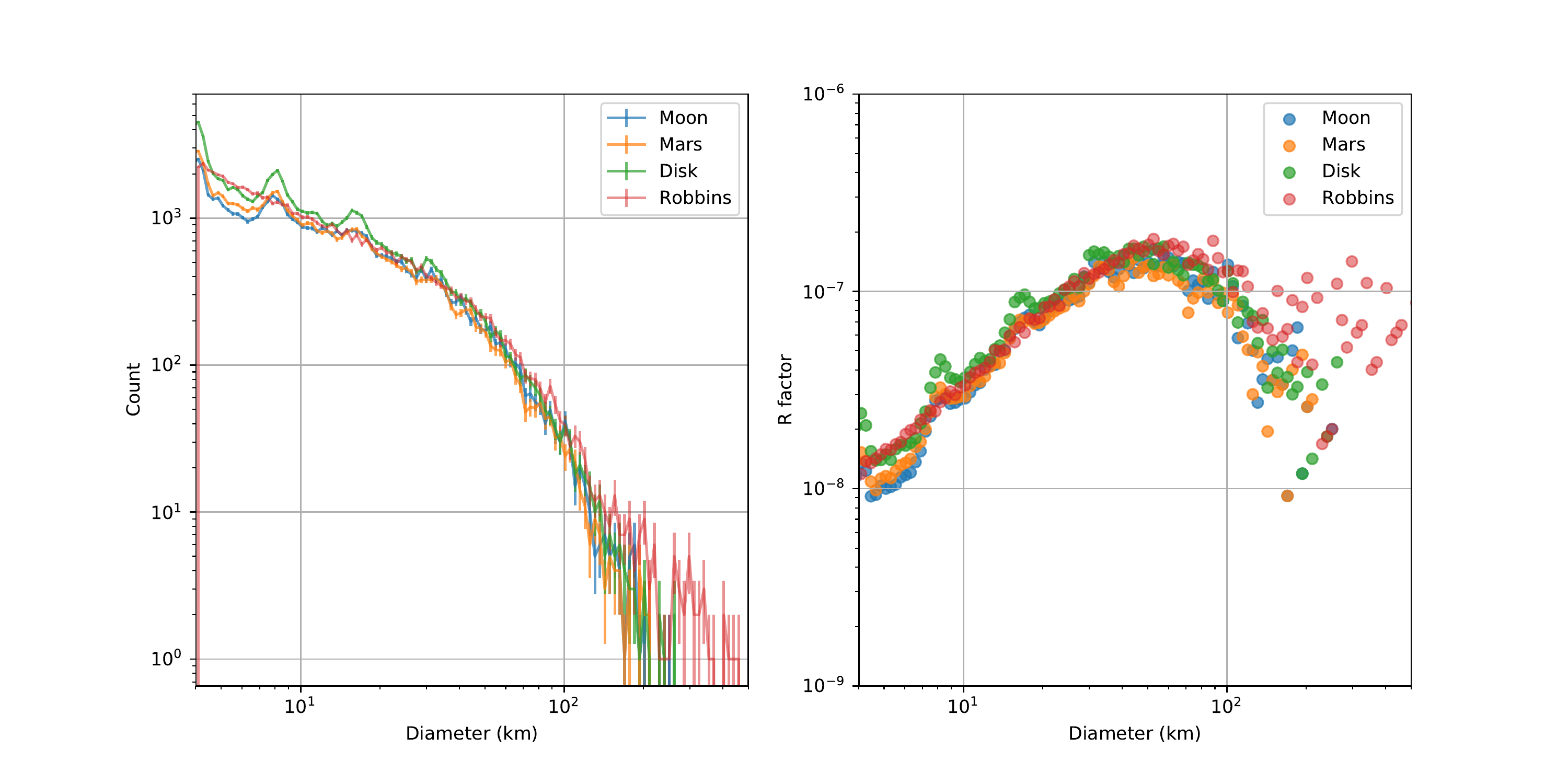}
	\caption{(left) Crater population as a function of crater diameter (km) for the datasets generated by the CDAs. (Right) R factor \citep{Arvidson1979} for the same dataset. The raw dataset from the CDA contains aliasing due to the small number of image resolutions used in the algorithm. This discretization has be removed from the data by applying a random jitter to the crater sizes equal to 5\%, smaller than the mean diameter error over all CDA datasets in table \ref{T:global}.}\label{F:power}
\end{figure*}

First, I compare the complete dataset generated by CDAs to the \citet{Robbins2012} dataset. Figure \ref{F:power} shows the crater distribution binned by diameter following the power law distribution used in \citet{Robbins2012}, and shows good agreement between the CDAs and the expected power law distribution. I used 16 bins per octave \citep{Robbins2012} of crater size instead of the 2 bins used in \citep{Stepinski2009} and \citet{Arvidson1979}. The discretization present in the diameter measurements from the CDAs has been removed from the data by applying a Gaussian noise multiplier (with magnitude of 5\%, smaller than the global mean diameter error in table \ref{T:global} of 7\%) to each data point. With only 2 bins/octave \citep{Arvidson1979} the distributions would agree with each other without the need for the de--aliasing jitter in the CDA data. The peaks at 8km and 16km are residuals of this jittering process and represent the smallest crater sizes found in the most common image resolutions used in the experiments. The CDA finds 80\% of the craters larger than 10 km diameter listed in the \citet{Robbins2012} database, and 75\% of craters larger than 4km in diameter. Craters below 4km are omitted from this dataset because of the lack of DTM data that resolve these craters.

\begin{table*}[h]
	\centering
	\scalebox{0.7}
	{\begin{tabular}{|l|l||c|c|c|c|c|c|c|c|}
		\hline
		        &            & Apron & Basin & Highland & Impact & Lowland & Polar & Transition & Volcanic\\
		\hline
		Robbins	& Count   	 & 116	& 467	& 41,749 & 3,016 & 2,858 & 660	 & 3,587 & 5,112 \\
		\hline
		Mars	& Count		 & 128	& 585	& 40,181 & 3,129 & 3,322 & 911	 & 3,645 & 5,866 \\
				& TPR (\%)   & 49	& 53	& 79	 & 72	 & 68	 & 44	 & 68	 & 69	 \\
		\hline
		Moon	& Count 	 & 108	& 521	& 38,376 & 2,990 & 3,089 & 785	 & 3,397 & 5,473 \\
				& TPR (\%)   & 56	& 58	& 81	 & 76	 & 73	 & 52	 & 72	 & 74	 \\
		\hline
		Disk	& Count      & 218	& 1,008	& 48,328 & 4,020 & 5,314 & 1,389 & 5,716 & 9,740 \\
				& TPR	(\%) & 20	& 26	& 59	 & 52	 & 42	 & 26	 & 40	 & 39	 \\
		\hline
	\end{tabular}}
	\caption{Distribution of craters by geologic unit type given in \citet{Tanaka2014}. The `Robbins' row gives the distribution of craters derived from craters in the \citet{Robbins2012} database. The True Positive Rate (TPR) gives the percentage of craters found by the CDA that exist in the \citet{Robbins2012} database. }\label{T:unittype}
\end{table*}

\begin{figure}[htbp]
	\centering
	\includegraphics[keepaspectratio,width=3in]{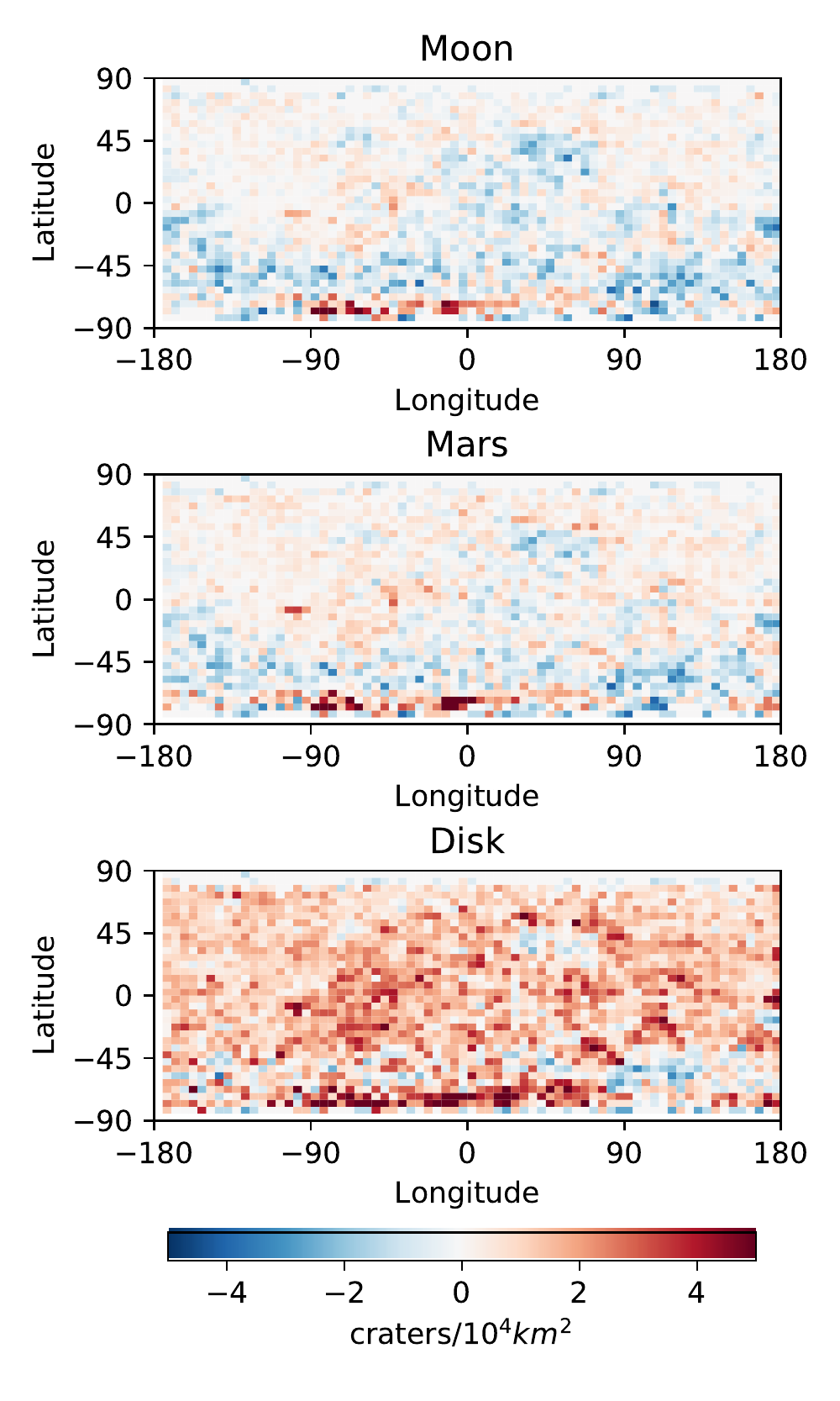}
	\caption{Plate Car\'ee maps of the crater number predictions from the CDA relative to the \citet{Robbins2012} dataset, binned into 5 degree square bins and scaled to represent the number of craters per 10,000 square kilometer predicted by each CDA in excess of the \citet{Robbins2012} database. Positive numbers (reds) represent an over--prediction by the CDA and negative numbers (blues) represent an under--prediction.}\label{F:craterloc2d}
\end{figure}

Table \ref{T:unittype} gives the crater numbers in each of the geologic unit types listed in \citet{Tanaka2014} for the 3 CDAs and the \citet{Robbins2012} database. The numbers are similar in the two ring--finding CDAs and the \citet{Robbins2012} database, although there are craters listed in the CDA datasets that are not present in the \citet{Robbins2012} database (the TPR percentage shown in the table reflects this). The disk--finding CDA tends to find many more craters in all geologic units and has more false positives (lower TPR) as a result. Figure \ref{F:craterloc2d} shows the same results but binned by longitude and latitude instead of geology. The two ring--finding CDAs tend to under--predict craters in regions with many craters, and over--predict in regions with few craters. The disk--finding CDA over--predicts the number of craters almost everywhere.

\subsection{Matched Craters}
\label{SS:matchedcraters}
\begin{figure*}[htbp]
	\centering
	\includegraphics[keepaspectratio,width=0.8\textwidth]{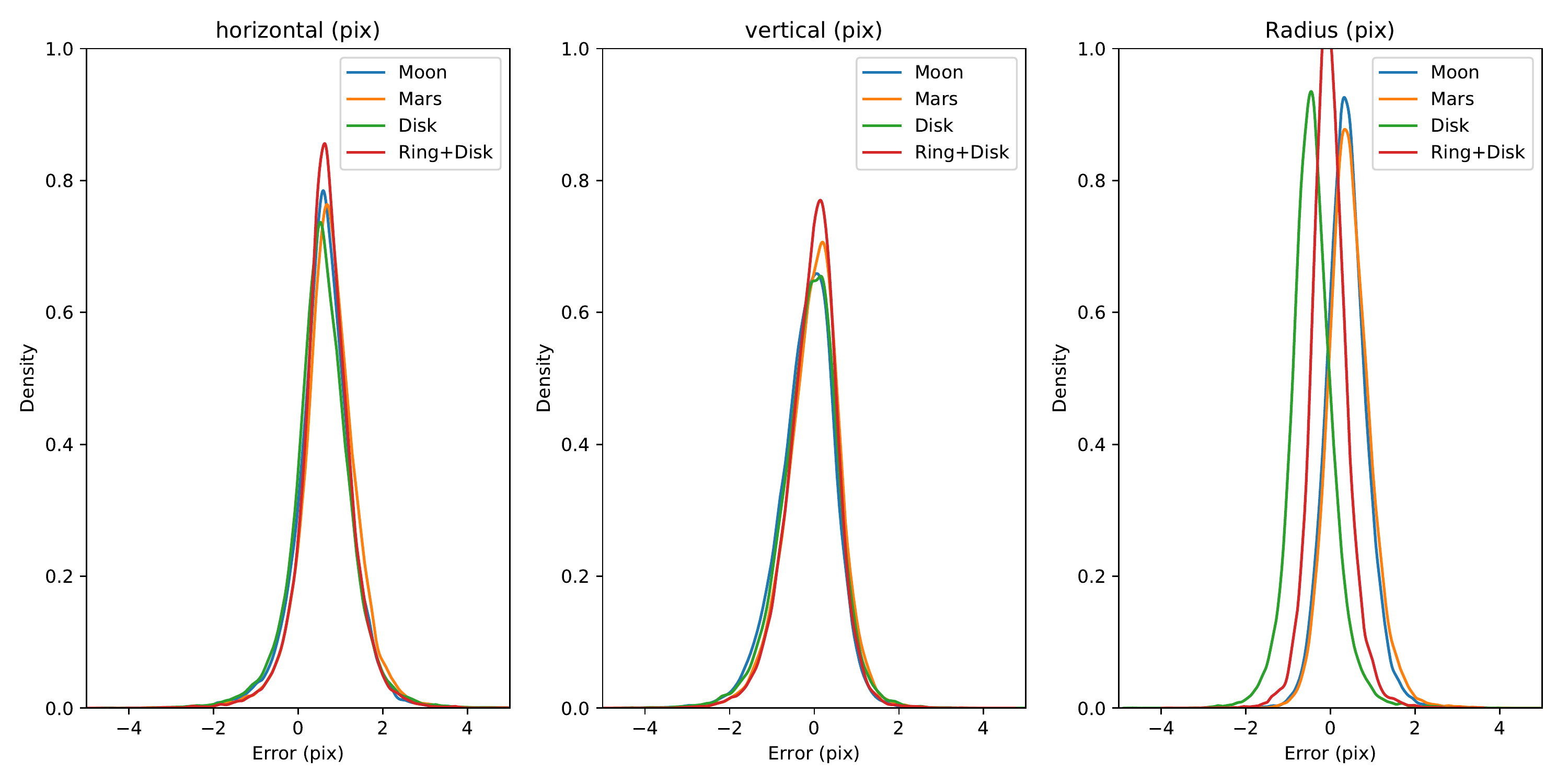}
	\caption{Distribution of pixel level differences between the CDA crater detection and the \citet{Robbins2012} dataset. Two additional merged CDA datasets are also included. ``Ring+Disk'' includes craters found in both the Mars CDA and the Disk CDA. Density is given in units of "per pixel" and is normalized.}\label{F:matched_pix_errors}
\end{figure*}

The \citet{Robbins2012} dataset used here contains 57,564 craters greater than 4km in diameter. The ring CDAs tested find 75\% of the craters in the \citet{Robbins2012} dataset with a median difference in location of 2\% and diameter of 5\% measured in geophysical units relative to the crater diameter. This difference is typical of variability between human analysts in crater studies \citep{Robbins2014}. In raw pixel terms, the differences in position and size between the CDA and Robbins datasets are typically 1 or 2 pixels. The distribution for each metric in pixel space is shown in figure \ref{F:matched_pix_errors}, with the median and inter-quartile ranges given in table \ref{T:pixelerrors}.

\begin{table*}[h]
	\centering
	\small
	\begin{tabular}{|c||c|c|c|c|}
		\hline
		&\textbf{Moon} &\textbf{Mars}&\textbf{Disk}&\textbf{Ring+Disk} \\
		\hline
		Horizontal (longitudinal) & $0.6\dpm{0.4}{0.4}$ & $0.7\dpm{0.4}{0.4}$ & $0.6\dpm{0.4}{0.4}$ & $0.7\dpm{0.3}{0.3}$ \\
		Vertical (latitudinal) & $-0.1\dpm{0.4}{0.4}$ & $0.03\dpm{0.4}{0.4}$ & $-0.06\dpm{0.5}{0.5}$ & $0.009\dpm{0.4}{0.3}$ \\
		Diameter & $0.4\dpm{0.3}{0.3}$ & $0.4\dpm{0.3}{0.3}$ & $-0.4\dpm{0.3}{0.3}$ & $-0.05\dpm{0.2}{0.3}$ \\
		\hline
	\end{tabular}
	\caption{Median and inter-quartile ranges for the image level differences between the \citet{Robbins2012} crater database and the CDA predictions. All values are given as median and interquartile values of the pixel level differences between the CDA and \citet{Robbins2012} data.}\label{T:pixelerrors}
\end{table*}

The ring trained CDAs and the disk trained CDA have similar accuracy on the location but the opposite sign in the crater diameter differences. This apparent bias may be a result of the method used to generate each prediction, as the disk--finding CNN uses a \citet{Sobel1968} transform to convert the predicted disks into rings, and places the ring within the perimeter of the disk, instead of on the outer edge.

This bias can be reduced by combining the results from the ring trained CDA and disk trained CDA such that only craters found by both CDAs are considered detections. This is shown as the ``Ring+Disk'' result in table \ref{T:pixelerrors} and figure \ref{F:matched_pix_errors}. The absolute mean difference in diameter between the CDA and \citet{Robbins2012} dataset decreases from 0.5 pixels to 0.05 pixels. The trade--off for this improved accuracy is that only craters found by both CDAs can be improved, and the recall of the worst CDA limits the number of craters that can be improved. In this dataset, 63\% of the existing craters are found by both CDAs providing a smaller catalog of more confidently detected craters.

\begin{figure*}
	\centering
	\includegraphics[keepaspectratio,width=0.8\textwidth]{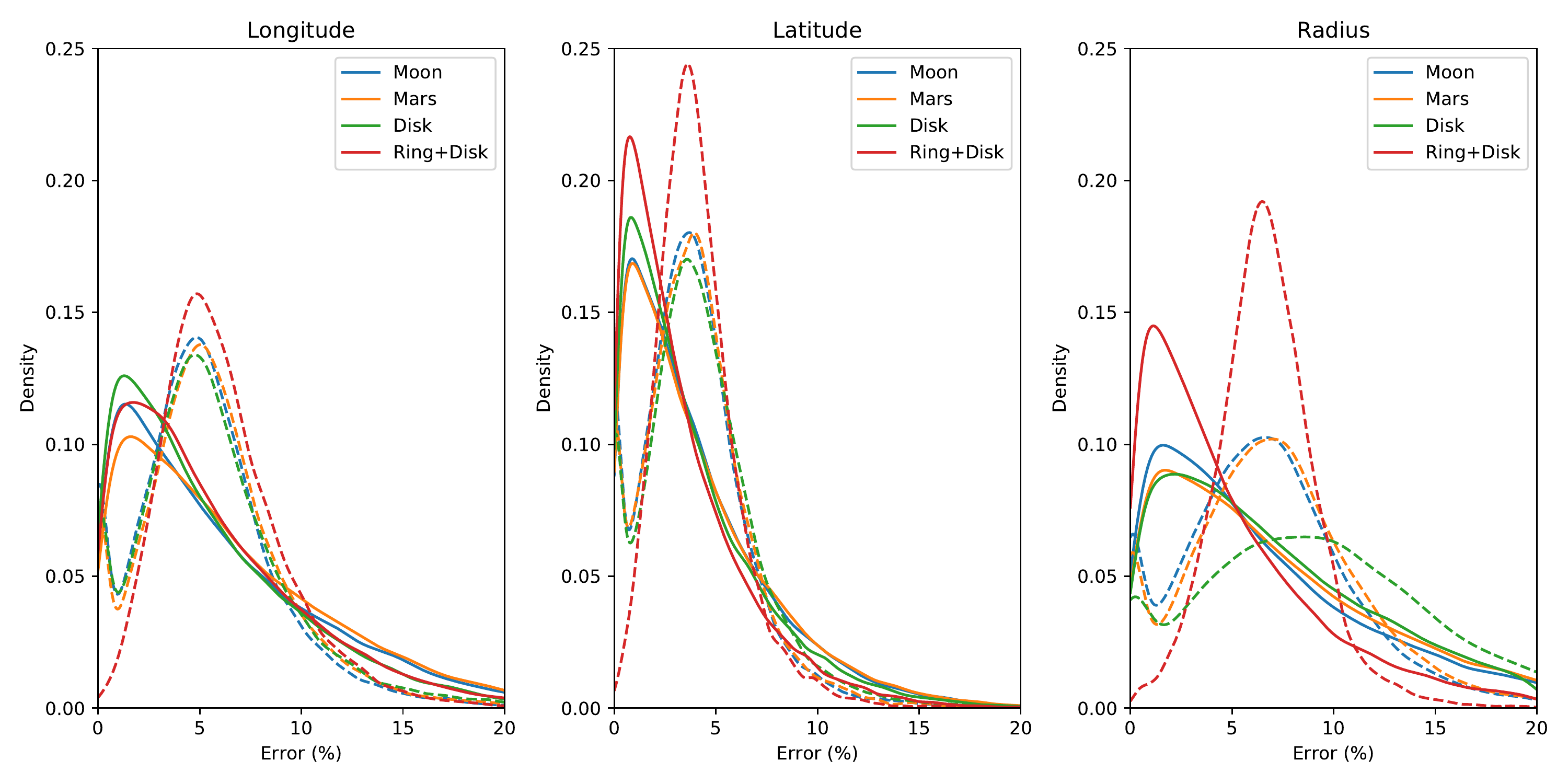}
	\caption{Error density plots for craters found by each CDA with a matching crater in the \citet{Robbins2012} dataset, given as the absolute fractional error relative to the crater diameter. (Left) Longitude errors, (center) latitude errors, (right) absolute diameter errors. Per--image statistics are shown with dashed lines, globally aggregated data is shown with solid lines The summary median and inter--quartile ranges for this data is given in table \ref{T:global}.}\label{F:matched_geo_errors}
\end{figure*}

In terms of geophysical location and size, the distributions of error in the longitude, latitude, and diameter of the matched craters are shown in figure \ref{F:matched_geo_errors}, with median and inter--quartile values given in table \ref{T:global} and \ref{T:validation}. After aggregating the per--image metrics to produce the global dataset, the CDA errors decrease as duplicate craters are filtered for higher precision crater location determined using the highest resolution image. 

In the global dataset size errors decrease from 6\% to 4\% (medians) in the combined ``Ring+Disk'' CDA, but the improvement comes at the expense of recall. In the globally aggregated data, the recall of the combined dataset is worse (at 60\%) than the recall of the worst individual CDA (the Disk CDA), while the precision is better (at 80\%) than the best CDA (the Mars CDA). The resulting $F_1$ score drops to 69\%, worse than the Mars Ring CDA and better than the Mars Disk CDA.


As a comparison with the errors shown here, \citet{Robbins2008} performed a similar study using human--derived datasets from MOLA DTMs and THEMIS imagery and noted that the DTM derived crater sizes are typically 1km larger than the imagery resolved counterparts. In this work the DTM derived crater sizes are 0.05km to 0.92km larger than the \citet{Robbins2012} data (25\% to 75\% percentiles) with the median crater being 0.44km larger. Twenty three percent of the DTM derived craters are smaller than their \citet{Robbins2012} counterpart.

\subsection{Missed Craters}
\label{SS:missedcraters}
None of the CDAs found every crater in the \citet{Robbins2012} list even if they found more than 57,564 craters in total. The missing craters don't need to share any characteristics but the \citet{Robbins2012} dataset includes a large number of parameters that might illustrate why some craters were missed. In particular, the \citet{Robbins2012} dataset contains the depths for each crater, including the depth relative to the crater edge (\texttt{DEPTH\_RIMFLOOR}), relative to the surrounding terrain (\texttt{DEPTH\_SURFFLOOR}), and the degradation/ preservation state (\texttt{DEGRADATION\_STATE}) that rates the condition of the crater from highly--degraded (1) to not--degraded (4). A `random decision forest' algorithm \citep{Ho1998} was used to identify these three parameters as most correlated with missing craters in this CDA relative to the \citet{Robbins2012}.

Comparing the Mars ring CDA with the \citet{Robbins2012} dataset (the other CDAs perform similarly), shallow craters are more likely to be missed than deep craters, and highly--degraded craters are more likely to be missed than non--degraded craters. For example craters with a rim-floor depth of 105m or less account for 15\% of the dataset, but accounted for 36\% of the missed craters. Highly degraded craters made up 45\% of all craters but 75\% of the missed craters (all other degradation states have a false negative rate of less than 5\%).  When combined, crater depth is a stronger determinant than the degradation state. In all degradation states, shallower craters were more likely to be missed than deep craters. In the worst case of highly degraded craters, shallower craters are missed at a rate 10 times higher than the deeper craters.

\begin{figure*}[htbp]
	\centering
	\includegraphics[keepaspectratio,width=0.8\textwidth]{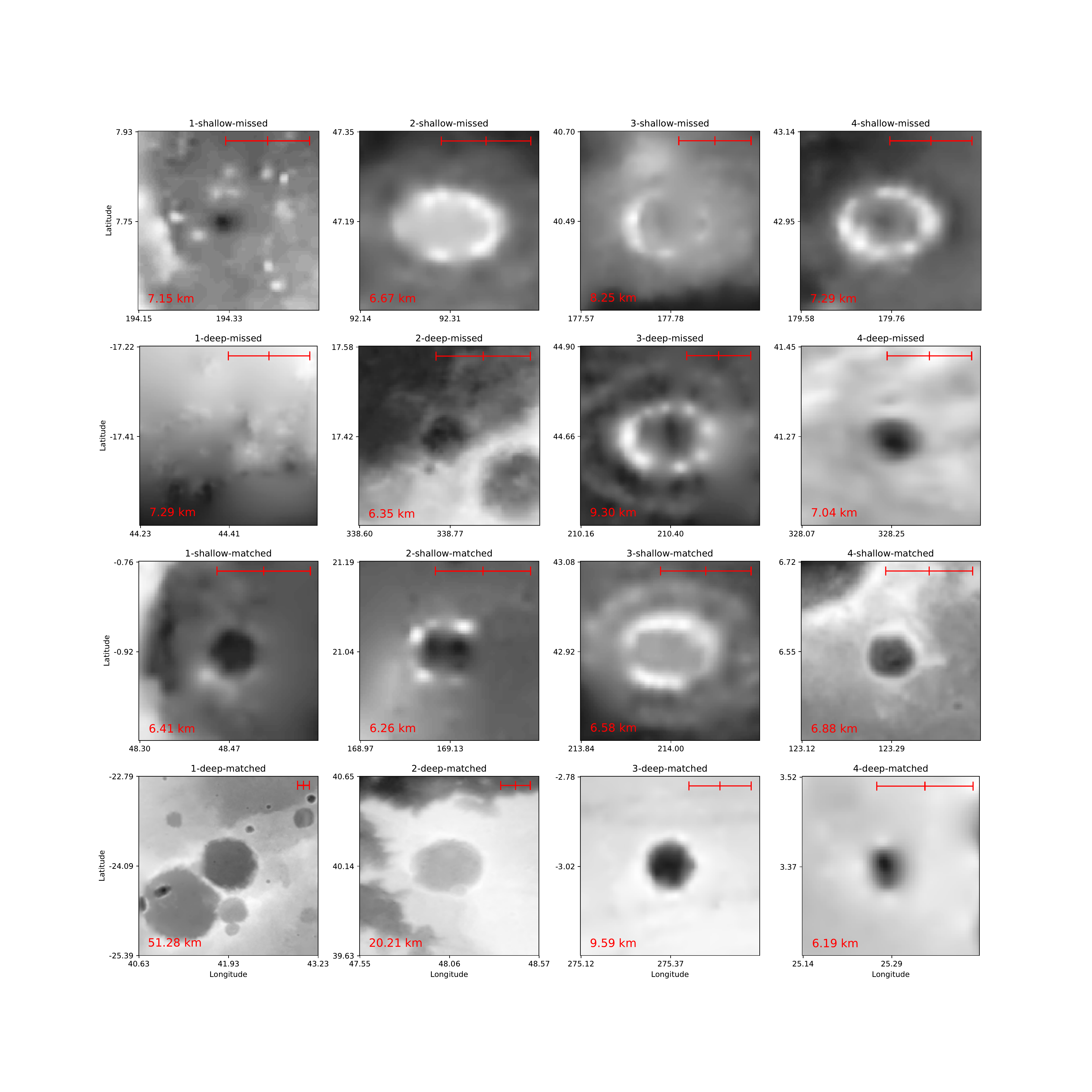}
	\caption{Randomly selected examples of craters from each degradation state (columns) and depth (alternating rows) that were missed (top two rows) or matched (bottom two rows). Each image includes the crater at the center of the image, and a border of 1 crater width on each side. The shades in each image are indicative of local topography in the image, but not necessarily the images presented to the CDA. The red text in bottom left of each panel gives the diameter of the centered crater. The scale bar at the top right is 10km in all images, marked at 0,5,10km.}\label{F:missed_examples}
\end{figure*}

Examples of missed and detected craters are shown in figure \ref{F:missed_examples}. Some of the less degraded craters can be found more easily in the THEMIS IR dataset used by \citet{Robbins2012} because of the contrasting effect of sunlight on the exposed edges of the crater.

Although the impact of the degradation state and crater depth were not known during the training step of this experiment, the different crater types were well represented in the crater populations used in training and validation datasets. If this were not the case, it might have been possible to improve the performance of the CNN on the shallow degraded craters by ensuring a representative sample of these craters in the training dataset.


\subsection{False Positives}
\label{SS:falsepositives}
The CDAs each detected craters that do not exist in the \citet{Robbins2012} dataset that are considered false positives. A large fraction of these detections were likely correctly identified as false positives (i.e., the craters do not exist), with a much smaller fraction being real craters missing from the \citet{Robbins2012} dataset.

Table \ref{T:unittype} gives the number of craters in each CDA and the \citet{Robbins2012} dataset, grouped by geologic type \citep{Tanaka2014}. The table also gives the \textit{true positive rate} or the fraction of craters in each CDA that correspond to a known crater. The remaining craters are the \textit{false positives}. The relatively poor performance of the CDAs in the \textit{Apron}, \textit{Basin}, and \textit{Polar} terrain only has a small impact on the overall results --- These terrains account for less than 1,500 craters i

\begin{figure*}[htbp]
	\centering
	\includegraphics[keepaspectratio,width=0.8\textwidth]{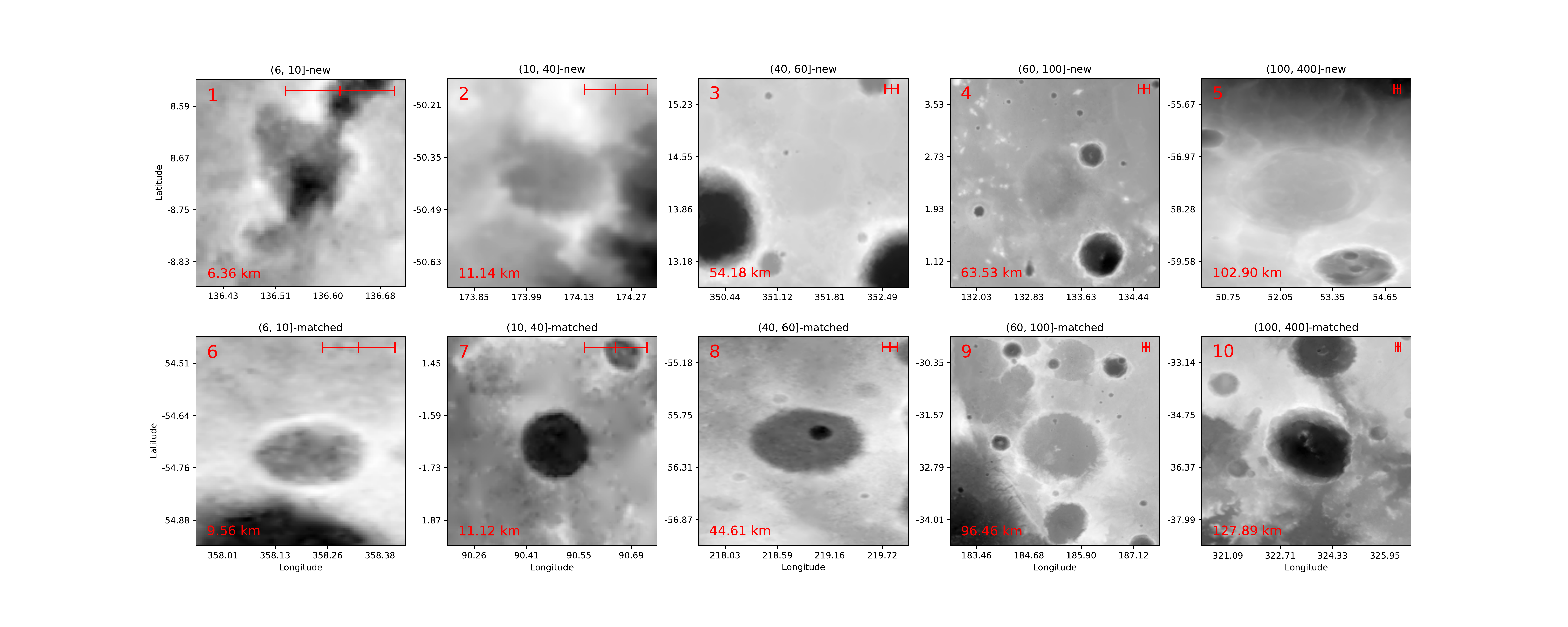}
	\caption{(top row) false positives in the Mars trained dataset, (bottom) true positives in the Mars trained dataset. The feature size increases from left to right (with the diameter range given in the title in kilometres) but the feature is randomly chosen from the CDA dataset. As in figure \ref{F:missed_examples} the identified crater is centered in the image with a 1 diameter border around it and red text in bottom left of each panel gives the diameter of the centered crater. The scale bar at the top right is 10km in all images, marked at 0,5,10km. (image 5 is Peneus Patera.)}\label{F:new_examples}
\end{figure*}

Examples of false positives in the Mars ring dataset are shown in figure \ref{F:new_examples}, grouped by the crater diameter. Some of the false positive detections have the appearance of craters while others are not obviously circular features (with 10,000+ false positives the small sample shown is random and not necessarily representative). For the larger detected features, many are paterae that are, correctly, not listed in the \citet{Robbins2012} crater catalog. Fifteen of the 20 largest diameter `false positives' correspond to mountains or paterae, and another 20 of the next largest 80 `false positive' detections are named features on Mars \textit{that are not craters}. The CDA is correct in identifying these circular features in the DTM, but incorrect in labelling them as craters.

For smaller sized features the results are less promising. A review of a random sample of 300 features below 5km in diameter did not identify any definitively new craters --- Approximately 30\% were depressions related to valleys or topography, but are not craters; 5\% were detections of craters with a diameter of 4km in the CDA but below this threshold in the \citet{Robbins2012} dataset (and are therefore removed from the dataset) ; 5\% of the craters are circular features in the DTM data, but disappear in higher resolution imagery. Most of the remaining 60\% are appropriately labelled as false positives and were not crater like even in the available DTM data. Only a small number of samples are possibly new craters, resulting in fewer than 100 new crater detections in the CDA datasets.

\citet{Silburt2019} attempted to answer a similar question by providing a sample of the false positives to researchers to categorize as crater or not. In that case 89\% were identified as new craters, in stark contrast to the numbers here. However, according to \citet{Robbins2012} their database is statistically complete below the lower limit of 4km considered here. For Lunar data, the crater catalog was less complete and 15\% of the new detections by the \citep{Silburt2019} CDA were below the lower limit of their ``ground truth'' database. Additionally, the test posed in \citet{Silburt2019} is framed differently, asking whether a human researcher would identify the feature as a crater, rather than asking whether the feature is actually a crater given all the available information.

\section{Conclusions}
\label{S:conclusions}
In this paper, I have applied a new Crater Detection Algorithm (CDA) to find craters in Mars digital terrain model. The CDA combines a multi--layer neural network to highlight circular features and a template correlation algorithm to determine their location and size. The best CDA used here finds 75\% of the craters listed in a comprehensive existing dataset \citep{Robbins2012}, in line with typical human performance on similar datasets \citep{Wetzler2007}. I also showed that a CDA trained on lunar data \citep{Silburt2019} performed well on the martian DTMs without further training.

The performance of each CDA was measured against the \citet{Robbins2012} crater list, and the predicted locations and sizes of craters compare well with that dataset. The \textit{ring} finding CDAs find craters over the entire martian surface with no significant bias in location, size, or geology, and with differences of around 5\% of the crater size and location relative to the \citet{Robbins2012} dataset, in line with estimated errors from human--generated crater datasets \citep{Robbins2014}. The \textit{disk} finding CDA performs worse in general and is significantly worse at finding craters in \textit{Impact}, \textit{Basin}, or \textit{Polar} geologic unit types than in the \textit{Highlands} \citep{Tanaka2014}.

The best CDA developed here misses many existing craters, and misidentifies other features as craters. The ring trained CDA missed 54\% of those craters in the most degraded state, and 80\% of those craters shallower than 105m from rim to floor. Given the large number of shallow craters missed, it might be possible to improve the performance of the CNN stage by increasing the `contrast' of the DTM images by limiting the vertical extent in each image, similar to the pre--processing technique used in \citet{Stepinski2009} to limit the horizontal scale of craters in each image.

A key feature of any automated CDA is the ability to make predictions rapidly and without human intervention. The CDA developed here can work with any standard DTM dataset and can generate 100--1,000 crater predictions per second on consumer hardware. DTMs generated from high resolution imagery can be used to generate catalogs of craters not available in current databases \citep{Lee2018a}, and could be incorporated into existing data processing pipelines. The overall speed benefit produced by the CDA depends on the number of manual corrections needed to create a final catalog. With approximately 25\% false positive rate and 25\% false negative rate (the Mars trained ring CDA), 1 in 4 proposed craters would be rejected by an expert human classifier, and 1 in 4 craters would need to be identified in the DTMs or by combining the catalog with other sources. Given the time required to accurately identify the rim of each crater in an image the CDA should significantly reduce the time taken to make a crater catalog. 

Finally, this work and others \citep{Silburt2019, Lee2018a} have shown that the CDA can be applied across different planets and DTM scales providing consistent datasets are available, allowing meaningful comparison between different planetary bodies using a consistent processing algorithm.

\section{Acknowledgments}
\label{S:acknowledgments}

Computations were performed on clusters and GPU enabled workstations at the University of Toronto. I am grateful to Ari Silburt and co-authors for making the source code for the DeepMoon CDA available. I thank Stuart Robbins and Christian Riedel for providing numerous comments and questions that have improved and clarified this work.

\section{Bibliography}
\bibliographystyle{abbrvnat}
\bibliography{library.bib}







\end{document}